\documentclass[a4paper,11pt]{article}
\usepackage[ams]{}
\usepackage{amsmath}
\usepackage{amsthm}
\usepackage{hyperref}
\usepackage{amssymb}
\usepackage{graphicx}
\usepackage[all,arc]{xy}
\usepackage[theorem]{}

\theoremstyle{plain}			

\newtheorem{thm}{Theorem}[section]

\newtheorem{cor}[thm]{Corollary}

\theoremstyle{definition}		

\begin{document}
\title{U-Duality and the Leech Lattice}
\author{Michael Rios\footnote{email: mrios@dyonicatech.com}\\\\\emph{Dyonica ICMQG}
\\\emph{Los Angeles, CA USA}  } \date{\today}\maketitle
\begin{abstract}
It has recently been shown that the full automorphism group of the Leech lattice, Conway's group $Co_0$, can be generated by $3\times 3$ matrices over the octonions.  We show such matrices are of type $F_4$ in $E_{6(-26)}$, the U-duality group for $\mathcal{N}=2$, $D=5$ exceptional magic supergravity. By mapping points of the Leech lattice to black hole charge vectors, it is seen $Co_0$ is generated by U-duality transformations acting as rotations in the charge space for BPS black holes. 
\\\\
$Keywords:$ Leech Lattice, U-duality, BPS black holes.
\end{abstract}

\newpage
\tableofcontents
\section{Introduction}
\indent Recently, Wilson has given a three-dimensional octonionic construction of the Leech lattice and has shown its automorphism group, Conway's group $Co_0=2\cdot Co_1$, is generated by $3\times 3$ matrices over the octonions \cite{1,2}.  Octonions have found wide application in the black hole-qudit correspondence \cite{3}-\cite{9}\cite{19,34,36,39,40}, through the study of the quantum informational interpretations of supergravities arising from toroidally compactified M-theory and magic supergravity theories \cite{10}-\cite{23},\cite{28}-\cite{30}\cite{33,37,38}.  The $\mathcal{N}=2$, $D=5$ exceptional magic supergravity, in particular, has the exceptional Jordan algebra over the octonions $J^{\mathbb{O}}_3$ as its charge space with U-duality group $E_{6(-26)}$ \cite{14,20}.

Building on the results of Wilson, we show the generators of the automorphism group of the Leech lattice are elements of $F_4$, expressed as unitary $3\times 3$ matrices over the octonions $\mathbb{O}$.  As $F_4\subset E_{6(-26)}$, the generators of the automorphisms of the Leech lattice can be interpreted as U-duality transformations in $\mathcal{N}=2$, $D=5$ exceptional magic supergravity.  We map points of the Leech lattice to elements of the exceptional Jordan algebra, where generators of the full automorphism group of the Leech lattice are rotations in the charge space of BPS black holes.

\section{Octonions, E8 and the Leech Lattice}

\subsection{The Octonions}
Let $V$ be a finite dimensional vector space over a field $\mathbb{F}=\mathbb{R},\mathbb{C}$.  An $algebra$ $structure$ on $V$ is a bilinear map
\begin{eqnarray}
V\times V \rightarrow V \\
(x,y)\mapsto x\bullet y.
\nonumber
\end{eqnarray}
A $composition$ $algebra$ is an algebra $\mathbb{A}=(V,\bullet)$, admitting an identity element, with a non-degenerate quadratic form $\eta$ satisfying
\begin{eqnarray}
\forall x,y\in\mathbb{A}\quad \eta(x\bullet y)=\eta(x)\eta(y).
\end{eqnarray}
If $\exists x\in\mathbb{A}$ such that $x\neq 0$ and $\eta(x)=0$, $\eta$ is said to be $isotropic$ and gives rise to a $split$ $composition$ $algebra$.  When $\forall x\in\mathbb{A}$, $x\neq 0$, $\eta(x)\neq 0$, $\eta$ is $anisotropic$ and yields a $composition$ $division$ $algebra$.
\begin{thm}
A finite dimensional vector space $V$ over $\mathbb{F}=\mathbb{R},\mathbb{C}$ can be endowed with a composition algebra structure if and only if $\textrm{dim}_{\mathbb{F}}(V)=1,2,4,8$.  If $\mathbb{F}=\mathbb{C}$, then for a given dimension all composition algebras are isomorphic.  For $\mathbb{F}=\mathbb{R}$ and $\textrm{dim}_{\mathbb{F}}(V)=8$ there are only two non-isomorphic composition algebras: the octonions $\mathbb{O}$ for which $\eta$ is anisotropic and the split-octonions $\mathbb{O}_s$ for which $\eta$ is isotropic and of signature $(4,4)$.  Moreover for all composition algebras, the quadratic form $\eta$ is uniquely defined by the algebra structure.
\end{thm}

The octonion composition division algebra $\mathbb{O}$ has an orthonormal basis $\{e_0=1,e_1,...,e_7\}$ labeled by the projective line $PL(7)=\{\infty\}\cup \mathbb{F}_7$, with product given by $e_1e_2=-e_2e_1=e_4$ and images under $e_t\mapsto e_{t+1}$ and $e_t\mapsto e_{2t}$ \cite{1,2}.  The norm is given by $\eta(x)=x\overline{x}$, where $\overline{x}$ is the conjugate of x.  

\subsection{$E_8$ and the Leech Lattice}
\indent The $E_8$ lattice embeds in the octonion algebra $\mathbb{O}$ in many different ways.  Following Wilson \cite{1,2}, we take the 240 roots to be 112 octonions $\pm e_t\pm e_u$ for any distinct $t,u\in PL(7)$, and 128 octonions of the form $\frac{1}{2}(\pm 1\pm e_1\pm \cdots \pm e_7)$ that have an odd number of minus signs.  The lattice spanned by these 240 octonions will be denoted as $L$, which is a scaled copy of the $E8$ lattice.

Using $L$ as a copy of $E8$ in the octonions, the \emph{octonionic Leech lattice} $\Lambda_{\mathbb{O}}$ is defined \cite{1,2} as the set of triples $(x,y,z)\in\mathbb{O}^3$, with norm\newline $N(x,y,z)=\frac{1}{2}(x\overline{x}+y\overline{y}+z\overline{z})$, such that\newline
\indent 1. $x,y,z\in L$\newline
\indent 2. $x+y, x+z, y+z\in L\overline{s}$\newline
\indent 3. $x+y+z\in Ls$\newline
Wilson has shown $\Lambda_{\mathbb{O}}$ is isometric to the Leech lattice \cite{2}.

\begin{thm}
(Wilson) The full automorphism group $Co_0$ of the Leech lattice is generated by the following symmetries \cite{1}:\newline
\indent (i) an $S_3$ of coordinate permutations;\newline
\indent (ii) the map $r_0:(x,y,z)\mapsto(x,ye_1,ze_1)$\newline
\indent (iii) the maps $x\mapsto \frac{1}{2}(x(1-e_1))(1+e_t)$ for $t=2,3,4,5,6,7$\newline
\indent (iv) the map $(x,y,z)\mapsto\frac{1}{2}((y+z)s,x\overline{s}-y+z,x\overline{s}+y-z)$
\end{thm}

\noindent In the proof of the theorem, to construct a non-monomial symmetry, Wilson takes $s$ to be a complex number $s=\frac{1}{2}(-1+\sqrt{-7})$, satisfying $s\overline{s}=2$ \cite{1}. Note: The maps given in $(iii)$ act on all three coordinates simultaneously.

\begin{cor}
The full automorphism group $Co_0$ of the Leech lattice is generated by $F_4$ symmetries.
\end{cor}
\noindent \begin{proof}The $S_3$ coordinate permutations of Wilson's theorem are given by six matrices in the usual $O(3)$ representation, where it is known $O(3)\subset F_4$ \cite{24}.  For the remaining symmetries, we note that $F_4$ contains all unitary $3\times 3$ matrices over the octonions \cite{24,30,32} and give the corresponding unitary matrices for each symmetry listed in Wilson's theorem.  The matrices are explicitly:\\

\noindent $(ii):\quad R_1=\left(\begin{array}{ccc} 1 & 0 & 0 \\ 0 & e_1 & 0 \\ 0 & 0 & e_1  \end{array}\right)$\\
$(iii):\quad R_2=\frac{1}{\sqrt{2}}\left(\begin{array}{ccc} 1-e_1 & 0 & 0 \\ 0 & 1-e_1 & 0 \\ 0 & 0 & 1-e_1  \end{array}\right),$\\
\indent$\quad\quad R_3=\frac{1}{\sqrt{2}}\left(\begin{array}{ccc} 1+e_t & 0 & 0 \\ 0 & 1+e_t & 0 \\ 0 & 0 & 1+e_t  \end{array}\right)$\\
$(iv):\quad R_4=\frac{1}{2}\left(\begin{array}{ccc} 0 & \overline{s} & \overline{s} \\ s & -1 & 1 \\ s & 1 & -1  \end{array}\right)$\\
and noting $e_1^2=e_t^2=-1$ and $s\overline{s}=2$, it is seen the matrices are unitary.\end{proof}
\section{Jordan Algebras and $D=5$ Magic Supergravity}
 
\subsection{Jordan Algebras}
\indent A Jordan algebra over a field $\mathbb{F}=\mathbb{R},\mathbb{C}$ is a vector space over $\mathbb{F}$ equipped with a bilinear form (Jordan product) $(X,Y)\rightarrow X\circ Y$ satisfying $\forall X,Y$:
\begin{displaymath}
X\circ Y = Y\circ X
\end{displaymath}
\begin{displaymath}
X\circ (Y\circ X^2)=(X\circ Y)\circ X^2.
\end{displaymath}
Given an associative algebra $\mathcal{A}$, we can define the Jordan product $\circ$ using the associative product in $\mathcal{A}$, given by $X\circ Y\equiv \frac{1}{2}(XY+YX)$.  Under the Jordan product, ($\mathcal{A},\circ$) becomes a \emph{special} Jordan algebra.  Any Jordan algebra that is simple and not special is called an \emph{exceptional} Jordan algebra or Albert Algebra.  The exceptional Jordan algebras are 27-dimensional Jordan algebras over $\mathbb{R}$, isomorphic to either $J^{\mathbb{O}}_3$ or $J^{\mathbb{O}_s}_3$.  The automorphism and reduced structure groups for these algebras are denoted as $\textrm{Aut}(J^{\mathbb{A}}_n)$ and $\textrm{Str}_0(J^{\mathbb{A}}_n)$, respectively.  In the case of $J^{\mathbb{O}}_3$, the Jordan algebra of $3\times 3$ Hermitian matrices over $\mathbb{O}$, $\textrm{Aut}(J^{\mathbb{O}}_3)=F_4$ and $\textrm{Str}_0(J^{\mathbb{O}}_3)=E_{6(-26)}$ \cite{14,20,24}.

The rank one elements of $J^{\mathbb{O}}_3$ are points of the octonionic projective plane $\mathbb{OP}^2=F_4/SO(9)$, also known as the \emph{Cayley plane}.   $F_4$ and $E_{6(-26)}$ act on the Cayley plane as isometry and collineation groups, respectively \cite{24,25,26,30}.

\subsection{$D=5$ Magic Supergravity}
\indent In $D=5$, the $\mathcal{N}=8$ and $\mathcal{N}=2$ magic supergravities are coupled to 6, 9, 15 and 27 vector fields with U-duality groups $SL(3,\mathbb{R})$, $SL(3,\mathbb{C})$, $SU^\ast(6)$, $E_{6(-26)}$ and $E_{6(6)}$, respectively \cite{4,13,14}.  The orbits of BPS black hole solutions were classified \cite{13,22} by studying the underlying Jordan algebras of degree three under the actions of their reduced structure groups, $\textrm{Str}_0(J_3^{\mathbb{A}})$, corresponding to the U-duality groups of the $\mathcal{N}=8$ and $\mathcal{N}=2$ supergravities.  This is seen by associating a given black hole solution with charges $q_I$ ($I=1,...,n_V)$ an element
\begin{equation}
J=\sum_{I=1}^{n}q^Ie_I=\left(\begin{array}{ccc} r_1 & A_1 & \overline{A}_2 \\ \overline{A}_1 & r_2 & A_3 \\ A_2 & \overline{A}_3 & r_3  \end{array}\right)\quad r_i\in\mathbb{R}, A_i\in\mathbb{A}
\end{equation}
of a Jordan algebra of degree three $J_3^{\mathbb{A}}$ over a composition algebra $\mathbb{A}$, where the $e_I$ form a basis for the $n_V$-dimensional Jordan algebra.  This establishes a correspondence between Jordan algebras of degree three and the charge spaces of BPS black holes \cite{13}.  The entropy of a black hole solution can be expressed \cite{4,13,23} in the form
\begin{equation}
S=\pi\sqrt{|I_3(J)|}
\end{equation}
where $I_3$ is the cubic invariant given by the determinant 
\begin{equation}
I_3=\textrm{det}(J).
\end{equation}
The U-duality orbits classify the black hole solutions via rank as
\begin{equation}\begin{array}{rcl}
\textrm{Rank}\thinspace J=3\quad \textrm{iff}\quad I_3(J)\neq 0\hspace{3pt}\qquad\quad\qquad S\neq 0,\thinspace \textrm{1/8-BPS}\hfill\\
\textrm{Rank}\thinspace J=2\quad \textrm{iff}\quad I_3(J)=0, J^{\natural} \neq 0\qquad S=0,\thinspace \textrm{1/4-BPS}\hfill\\
\textrm{Rank}\thinspace J=1\quad \textrm{iff}\quad J^{\natural} = 0,\thinspace\thinspace J\neq 0\thinspace\thinspace\quad\qquad S=0,\thinspace\textrm{1/2-BPS}\hfill
\end{array}
\end{equation}
and $J^{\natural}$ is the quadratic adjoint map given by
\begin{equation}
J^{\natural}=J\times J = J^2-\textrm{tr}(J)J+\frac{1}{2}(\textrm{tr}(J)^2-\textrm{tr}(J^2))I.
\end{equation}
The U-duality group $E_{6(-26)}$ preserves the cubic invariant and thus the entropy of BPS black holes in $\mathcal{N}=2$, $D=5$ exceptional magic supergravity.

\section{The Leech Lattice and U-Duality}

\subsection{The Leech Lattice and BPS Black Holes}
Recall that Wilson's octonionic construction of the Leech lattice uses triples $v=(x,y,z)\in\mathbb{O}^3$, where $x,y,z\in L$, with $L$ being a copy of the $E_8$ root lattice.  We can thus map such triples to elements of the exceptional Jordan algebra $J^{\mathbb{O}}_3$ via:
\begin{equation}
v\mapsto \overline{v}v^{T}=V\in J^{\mathbb{O}}_3
\end{equation}
This allows one to map points of the Leech lattice to black hole charge vectors in $\mathcal{N}=2$, $D=5$ exceptional magic supergravity.  Denoting the set of all such mapped points as $J^{\mathbb{O}}_3(\Lambda)$ we recover a map from the (octonionic) Leech lattice to the exceptional Jordan algebra:
\begin{equation}
\Lambda_{\mathbb{O}}\rightarrow J^{\mathbb{O}}_3(\Lambda)
\end{equation}
We shall refer to the black hole solutions with charge vectors in $J^{\mathbb{O}}_3(\Lambda)$ as the \emph{Leech BPS black holes}, which in general admit charges in a 24-dimensional charge space since they are constructed from elements of $\mathbb{O}^3$. 
\subsection{Conway's Group $Co_0$ and U-Duality}
Given a charge vector for a Leech BPS black hole, we can act on it via the $F_4$ generators of the Conway group $Co_0$ as:
\begin{equation}
V\mapsto UVU^{\dagger}=V'
\end{equation}
where $UVU^{\dagger}$ is unambiguous since each $U$ contains only a single octonion.

Wilson has given an octonionic description of the 196560 (norm four) vectors that span the Leech lattice \cite{2}.  They are given by:\\
\indent $(2\lambda,0,0)$\;\;\qquad\qquad\qquad\qquad\qquad\qquad($3\times 240=720$)\newline
\indent $(\lambda\overline{s},\pm(\lambda\overline{s})j,0)$\qquad\qquad\qquad\qquad\qquad($3\times 240\times 16=11520$)\newline
\indent $((\lambda s)j,\pm\lambda k,\pm(\lambda j)k)$\;\quad\qquad\qquad\qquad($3\times 240\times 16\times 16=184320$)\newline
where $\lambda$ is a root of the $E_8$ lattice and $j,k\in\{\pm e_t|t\in PL(7) \}$ and all permutations of the three coordinates are allowed.

Mapping the 196560 norm four vectors of the first shell of the Leech lattice to rank one elements of $J^{\mathbb{O}}_3$ corresponds to mapping to 1/2-BPS black hole charge vectors.  Recall that geometrically such charge vectors are points of the Cayley plane $\mathbb{OP}^2$, so the generators of the Conway group $Co_0$, which are $3\times 3$ unitary matrices over the octonions (hence elements of $F_4$) give isometries of these 1/2-BPS black hole charge vectors in $\mathbb{OP}^2$.  As $F_4\subset E_{6(-26)}$, the isometries are U-duality rotations preserving the vanishing entropy of the corresponding 1/2-BPS black hole solutions.  

As $Co_0$ is the double cover of Conway's group $Co_1$, and $2^{1+24}\cdot Co_1$ is a maximal subgroup of the Monster, it is natural to seek a U-duality realization of the Monster group.  The Monster is known to act as the group of linear transformations on a vector space of dimension 196883+1, 98280=196560/2 dimensions of which come from the set of norm four vectors of the first shell of the Leech lattice, with opposite pairs identified. \cite{45}

\section{Conclusion}
Building on the results of Wilson \cite{1,2}, we have shown the automorphism group of the Leech lattice, $Co_0$, is generated by unitary $3\times 3$ matrices over the octonions, and are therefore $F_4$ transformations.  As $F_4\subset E_{6(-26)}$, and $E_{6(-26)}$ is the U-duality group of $\mathcal{N}=2$, $D=5$ exceptional magic supergravity, we interpreted such generators of $Co_0$ as U-duality transformations.  As the charge space for BPS black holes in $\mathcal{N}=2$, $D=5$ exceptional magic supergravity is the exceptional Jordan algebra $J^{\mathbb{O}}_3$, this shows $Co_0=2\cdot Co_1$ encodes U-duality transformations. Since $\textrm{Aut}(J^{\mathbb{O}}_3)=F_4$ and $\textrm{Str}_0(J^{\mathbb{O}}_3)=E_{6(-26)}$, and $F_4$ is the trace preserving part of $E_{6(-26)}$, the $F_4$ generators for $Co_0$ were shown to preserve the Leech lattice norm in $J^{\mathbb{O}}_3$, as expected from isometries.  

In other studies, the Leech lattice and the subgroup $M_{24}$ of its automorphism group have found applications in string theory \cite{41,42,43,44}, shedding light on monstrous moonshine \cite{45} and its generalizations.  It was found \cite{41} there is a \emph{Mathieu moonshine} relating the elliptic genus of $K3$ to the sporadic group $M_{24}$.  More recently, Mathieu moonshine was shown to hold for $D=4$, $\mathcal{N}=2$ string compactifications, which arise from heterotic strings on $K3\times T^2$ or from type II strings on Calabi-Yau threefolds, where the one-loop prepotential universally exhibits a structure encoding degeneracies of $M_{24}$ \cite{42}.

In moving from $D=5$ to $D=4$, the octonionic $\mathcal{N}=2$ magic supergravity exhibits an $E_{7(-25)}$ U-duality symmetry, containing a maximal subgroup $SO(2,10)\times SU(1,1)$ found in the moduli space of complex structure deformations of the FHSV model \cite{46}, which is directly related to heterotic strings on $K3 \times T^2$.  In \cite{47} it was argued the $D=4$, $\mathcal{N}=2$ octonionic magic supergravity admits a stringy interpretation closely related to the FHSV model. 

In $D=11$, Ramond noticed \cite{48,49} the massless supermultiplet of eleven-dimensional supergravity can be generated from the decomposition of certain representation of $F_4$ into those of its maximal compact subgroup $Spin(9)$.  It was argued \cite{49} that the Cayley plane, $\mathbb{OP}^2=F_4/SO(9)$, may geometrically shed light on the fundamental degrees of freedom underlying M-theory.  Building on these observations, Sati proposed \cite{50,51} a Kaluza-Klein construction, where hidden $\mathbb{OP}^2$ bundles with structure group $F_4$ encode the massless fields of M-theory, resulting in a 27-dimensional candidate for bosonic M-theory \cite{52}.  Perhaps $\mathcal{N}=2$, $D=5$ exceptional magic supergravity has its full intepretation in such a 27-dimensional model.  Moreover, the appearance of $F_4$ in a model of M-theory with Cayley plane fibers would allow a Leech lattice construction, with $F_4$ generators of $Co_0$ acting on such fibers.  It is a pressing question as to the role of the Monster group in 27-dimensions, as the 196560 norm four vectors of the Leech lattice in 24-dimensions contribute to the degrees of freedom of the Griess algebra on which the Monster acts.

\end{document}